\documentclass{svjour3}                     
\smartqed  
\usepackage{graphicx}
\usepackage{amsmath}
\usepackage{amssymb}
\usepackage{natbib}
\usepackage{color}
\usepackage{enumitem}

\newcommand{\appxlabel}[1]{\label{sec:#1}}        
\newcommand{\appx}[1]{Appendix~\ref{sec:#1}}       
\newcommand{\eqlabel}[1]{\label{eq:#1}}           
\newcommand{\eq}[1]{(\ref{eq:#1})}                
\def\eqtwo(#1,#2){(\ref{eq:#1},\ref{eq:#2})}      
\def\eqset(#1-#2){(\ref{eq:#1}--\ref{eq:#2})}     
\newcommand{\fig}[1]{fig.~\ref{fig:#1}}           
\newcommand{\myfigure}[4]{\begin{figure*}[htb]
#1
\caption[#2]{#2 #3}
\label{fig:#4}
\end{figure*}}
\newcommand{\secnlabel}[1]{\label{sec:#1}}         
\newcommand{\secn}[1]{Section~\ref{sec:#1}}        
\newcommand{\Secn}[1]{Section~\ref{sec:#1}}        

\renewcommand{\@}{\partial}
\newcommand{\Binom}[2]{\left(\begin{array}{c}#1\\#2\end{array}\right)}
\newcommand{\bydef}{\triangleq}
\newcommand{\const}{\mathrm{const}}
\renewcommand{\d}{\mathrm{d}}
\newcommand{\Df}[2]{\frac{\d #1}{\d #2}}

\newcommand{\df}[2]{\frac{\partial #1}{\partial #2}}
\newcommand{\ddf}[2]{\frac{\partial^2 #1}{\partial #2^2}}
\newcommand{\e}{\mathrm{e}} 
\newcommand{\infint}{\int\limits_{-\infty}^{\infty}}
\newcommand{\Real}{\mathbb{R}}

\newcommand{\+}[2]{\def#1{#2}}
\newcommand{\1}[2]{\renewcommand{#1}[1]{#2}}
\+{\A}{A}           
\+{\a}{a}           
\+{\alp}{\alpha}    
\+{\B}{B}           
\+{\bet}{\beta}     
\+{\Ccoeff}{C}      
\+{\C}{C}           
\+{\Ctilde}{\tilde{\Ccoeff}} 
\+{\Dcoeff}{D}      
\+{\D}{D}           
\+{\Dtilde}{\tilde{\Dcoeff}} 
\+{\del}{\delta}    
\+{\dirac}{\delta}  
\+{\est}{\mathrm{e.s.t.}} 
\+{\expdec}{\lambda}
\+{\E}{E}           
\+{\Fit}{R}         
\+{\Flux}{J}        
\+{\Ftilde}{\tilde{\Flux}} 
\+{\fit}{r}         
\+{\fitm}{\bar\fit} 
\+{\fitz}{\fit_0}   
\+{\fitg}{\fit_1}   
\+{\ftilde}{\tilde\fit} 
\+{\gDmax}{G}       
\+{\gam}{\gamma}    
\+{\h}{h}           
\+{\Integral}{I}    
\+{\Intv}{\tilde\Integral}    
\+{\j}{j}           
\+{\K}{K}           
\+{\Kone}{K_1}      
\+{\Ktwo}{K_2}      
\+{\k}{k}           
\+{\kurt}{\kappa}   
\+{\kurti}{\kurt_0} 
\+{\l}{\ell}        
\+{\Moment}{M}      
\+{\Momv}{\tilde\Moment}    
\+{\Mut}{m}         
\+{\m}{m}           
\+{\mean}{\mu}      
\+{\mbar}{\bar\mean}
\+{\mut}{D}         
\+{\n}{n}           
\+{\N}{N}           
\1{\O}{\mathcal{O}\left(#1\right)} 
\+{\PDF}{\pdf_*}    
\+{\P}{P}           
\+{\Prey}{R}        
\+{\pdf}{p}         
\+{\pdfini}{\pdf_0} 
\+{\Qconst}{Q}      
\+{\q}{q}           
\+{\Shift}{\Delta}  
\+{\s}{s}           
\+{\skew}{\gamma}   
\+{\skewi}{\gamma_0}   
\+{\Stddev}{\Sigma} 
\+{\stddev}{\sigma} 
\+{\stddevi}{\sigma_0} 
\+{\sbar}{\bar\stddev}
\+{\t}{t}           
\+{\tf}{\tau}       
\+{\ts}{\tau}       
\+{\th}{t}          
\+{\u}{u}           
\+{\V}{V}           
\+{\v}{v}           
\+{\vini}{\pdfini}  
\+{\vb}{\phi}       
\+{\VB}{\Phi}       
\+{\W}{W}           
\+{\w}{w}           
\+{\x}{x}           
\+{\xc}{x_c}        
\+{\xm}{\x_{-}}     
\+{\xp}{\x_{+}}     
\+{\xf}{\xi}        
\+{\xs}{\eta}       
\+{\y}{y}           
\+{\z}{z}           
\+{\zoo}{z}         

\journalname{Journal of Mathematical Biology}

\begin{document}
\title{A simple mathematical model of gradual Darwinian evolution:
Emergence of a Gaussian
trait distribution in adaptation along a fitness gradient
}
\author{Vadim N. Biktashev}
\institute{
  College of Engineering, Mathematics and Physical Sciences, University of Exeter\\
  Harrison Building, North Park Road, Exeter EX4 4QF, UK\\
}
\date{\today}

\maketitle 

\begin{abstract}
  We consider a simple mathematical model of gradual Darwinian
  evolution in continuous time and continuous trait space, due to
  intraspecific competition for common resource in an asexually
  reproducing population in constant environment, while far from
  evolutionary stable equilibrium.  The model admits exact analytical
  solution. In particular, Gaussian distribution of the trait emerges
  from generic initial conditions.
\end{abstract}

\section{Introduction}
\secnlabel{introduction}

The question considered in this paper is: suppose a population evolves
according to the Darwin's mechanism involving mutations and natural
selection, and some of its quantitative traits change gradually, what
is the rate of this gradual change? This question may be not the most
interesting when applied to analysis of past evolution, say fossil
records, where the epochs of such gradual changes are relatively short
compared to much longer epoch when the species appear unchanged
\citep[``punctuated equilibrium'',][]{Eldredge-Gould-1972}. However,
the speed of evolution is crucial in constant competition of taxa
(``Red Queen'' hypothesis, \citealt{Liow-etal-2011}; ``evolutional
arms race'', \citealt{Dawkins-Krebs-1979}), and is of considerable
practical importance in relation to present day phenomena such as
adaptation of pathogens to existing methods of treatment, or
adaptation of endangered species to changing environmental conditions.

The literature dedicated to this subject is vast. Here we mention 
only some cornerstone publications, most relevant to the present
communication, to designate its context and motivation.

Quantitative approach to evolution dates back at least to
Fisher's~\citeyearpar{Fisher-1930} book, which contained his famous
``Fundamental Theorem of Natural Selection'', stating that the rate of
increase of the mean fitness of a population at any moment of time,
attributable to to natural selection, equals the genetic variance of
fitness of that population at that moment of time. This result is as
elegant and powerful as it is difficult to apply correctly, since
its deceptively simple words encrypt
complicated concepts, as it took nearly 40 years to figure
out~\citep{Price-1972}. The next question is, of course, what
determines this variance in the population fitness, and how to predict
it. 

\emph{Adaptive dynamics} is
mainly concerned with qualitative questions such as direction of
evolution, stability of evolutionary steady states and speciation due
to branching~\citep{Geritz-etal-1998,Bowers-2011}. On the quantitative level, the
fundamental for adaptive dynamics is the ``canonical equation''. The
influential paper by \citet{Dieckmann-Law-1996} gives this equation in
the form (for a single selected trait)
\begin{equation}
  \Df{\mean}{\t} = \K(\mean)\,\left.\df{\fit(\x,\mean)}{\x}\right|_{\x=\mean}
                                                            \eqlabel{canonical}
\end{equation}
where $\mean(\t)$ is the average value of the trait at time $\t$,
$\fit(\x,\mean)$ measures fitness of individuals with trait value $\x$
in the environment of resitent trait values $\mean$ and the
coefficient $\K(\mean)$ is described as a ``non-negative coefficient,
\dots that scales the rate of evolutionary
change''. \citeauthor{Dieckmann-Law-1996} have endeavoured to derive
this coefficient based on a certain miscroscopic model of mutation and
selection processes, and come to the result that it is equal to the
variance of the population with respect to the quantiative trait,
which, when applied to the fitness, exactly reproduces the Fisher's
result. In their derivation, Dieckmann and Law made an essential
assumption, with reference to separation of times between mutations
and selection in the limit of slow evolution and the competitive
exclusion principle, that at each moment of time, the selection
reduces the population to a certain type, which however changes in
time due to random mutations (``quasi-monomorphic framework'').
This framework, under the name of the ``Trait Substitution Sequence''
model, has been rigorously justified by \citet{Champagnat-2006}, using
a stochastic model, under certain asymptotic assumption about the
mutation rate. Slightly simplifying, the key assumption is that
mutations are so rare that for a given population size, there is
sufficient time between consecutive mutations for the whole population
to convert to the new trait value if it is fitter than the
previous. They also consider the opposite limit which they call
``large-population limit with accelerated births and deaths'', in
which the population is so big and mutations so frequent that at any
time the population consists of many different types. This leads
to a deterministic model in the form of an integro-differential
equation, which is akin to Fisher's reaction-diffusion equation, only
for population distribution in the trait space, and is a
time-dependent variation of the~\citet{Kimura-1965} model.  Such
deterministic models have been studied in many works, %
e.g.~\citet{Calsina-Perello-1995,Gudelj-etal-2006,Schuster-2011}.
These works typically concentrate on the analysis of stationary
solutions corresponding to the evolutionary stable states, rather than
quantifying the speed with which these states may be approached.

\emph{Quantitative genetics} has developed a number of its own
approaches and investigated a great number of complicated problems
related to quantitative description of evolution.
  One approach is through the method of moments.
  An example is the paper by \citet{Barton-Turelli-1987} which
  considered multilocus determination of a quantitative trait in a
  sexually reproducing population, and in particular presented an
  infinite chain of ordinary differential equations for the moments of
  allelic distribution.
  A more recent example is a paper by \citet{Sasaki-Dieckmann-2011},
  which looks at multiple-peak character distributions, called
  ``olygomorphs'', whereas each of the morphs is treated by the method
  of moments.
The chain of equations for the moments is typically treated using a
``closure'' procedure, say by assuming that the distribution is
Gaussian. Gaussian distributions naturally occurred in a number of
studies, e.g. by \citet{Kimura-1965}, \citet{Lande-1975} and
\citet{Turelli-1984}, as stationary solution at stabilizing selection.
However, stabilizing selection is hardly relevant to description of
intermediate states of continuing evolution; hence as far as we can
see, the question whether and when Gaussian distribution may actually
realize during such evolution remains open.

Here we aim to analyze the speed of evolution while far from any
evolutionary stable state, based on the simplest possible meaningful
model. This is a deterministic integro-partial differential equation,
which is similar to various forms postulated or derived elsewhere.  We
also provide a simple derivation of this model from ``first
principles'', avoiding to make non-verifiable assumptions, for fear
that the ultimate results may become artefacts of any such
assumptions. We consider single asexually reproducing species, stick
with phenotypic description and use dynamic formalism, leaving all
stochastics within the underlying population dynamics model of
intraspecific competition. As a model of intraspecific competition, we
consider a predator-prey system where various predator populations
depend on a common prey and do not interact otherwise. All these
assumptions are admittedly rather restrictive; we believe that
\textit{ab initio} approach not only helps to make clear what
postulates the ultimate results depend on, but can also suggest a
basis for subsequent generalizations for more realistic
assumptions. The simplicity of the resulting mathematical model allows
its exhaustive rigorous study. In particular, the Gaussian shape of
the trait distribution does indeed emerge spontaneously. The practical
utility of the model is illustrated by providing treatment of a more
realistic model via asymptotic methods.

The structure of the paper is as follows. \Secn{derivation} introduces
the main equation and its variations. \Secn{normal} describes the
``normal solution'' of the simplified version of the model, which
underlies subsequent analysis. \Secn{general} goes on to consider the
general solution of the simplified model. These results are extended
to the more generic version of the model by means of a perturbation
theory in \secn{perturbation}. \Secn{discussion} is dedicated to the
discussion of the results. We conclude with \appx{AppA} with the
derivation of our model ``from the first principles'' and \appx{AppB}
with the proof of the theorem presented in \secn{general}.

\section{The model}
\secnlabel{derivation}

We consider
  diffusive, mutational spread of a phenotypic trait distribution during a
  transient phase of optimizing evolution, described by a
  deterministic model of the form
\begin{equation}
   \df{\u}{\t} = \left( 
    \fit(\x) - \N(\t)
  \right)  \u 
  + \df{}{\x} \left[
    \Ccoeff(\x) \u
    + \Dcoeff(\x) \df{\u}{\x}
  \right], 
  \quad
   \N(\t)  =\infint  \u(\x,\t)\,\d{\x},  \eqlabel{intdiffstart}
\end{equation}
where $\x$ is a continuous trait, $\u(\x,\t)\ge0$ is population
density in the trait space so that $\N(\t)>0$ is the total population
at time $\t$, $\fit(\x)$ is the fitness of trait value $\x$ measured
as the low-density reproduction rate of the given type and
corresponding carrying capacity of the habitat, and $\Ccoeff(\x)$ and
$\Dcoeff(\x)>0$ represent mutations, with $\Ccoeff(\x)$ for the
directed component and $\Dcoeff(\x)$ for the diffusive component. The
case $\Ccoeff(\x)=\Dcoeff'(\x)$ of this equation can be obtained as a
special case of deterministic integro-partial differential
equation~(4.5) derived in~\cite{Champagnat-etal-2006} as a weak limit
of a stochastic model, with convolution kernels $U=\const$ and
$V=\const$ (the ``mean-field'' case), and with an appropriate choice
of functions $b$ and $d$.  A simple non-rigorous derivation
of~\eq{intdiffstart} through a continuous-trait limit of a
deterministic population dynamics model is given in \appx{AppA}. The
model is admittedly rather simplified and ignores many important
evolutionary factors, e.g. frequency dependent selection.  Some
equivalent forms of special cases of this equation found in literature
will be mentioned below.

We shall first look at solutions of \eq{intdiffstart} in simplifying
assumptions regarding its coefficients, and then
  relax those assumptions by means of a perturbation theory.  So, in
the simplified version, we take that $\Dcoeff(\x)=\mut=\const$
(which is a significant limitation as normally one would
  expect that mutation rate is proportional to birth rate so should be
  varying together with the reproduction rate $\fit(\x)$).
Further, the coefficient $\Ccoeff(\x)$ represents possible mutations
bias with respect to the selected trait.  In many studies it is
assumed to be zero, however it may correspond to non-selective
evolution, discussed e.g. by~\citet{Koonin-2009}.  For the sake of
simplicity, we take $\Ccoeff=\const$; then without loss of generality
we take $\Ccoeff=0$, as a nonzero constant $\Ccoeff$ would simply add
$-\Ccoeff$ to the trait change rate.  So,
\begin{eqnarray}
  \df{\u}{\t} = \left( 
    \fit(\x) - \N(\t)
  \right)  \u 
  + \mut \ddf{\u}{\x} , 
  \qquad
  \N(\t)  =\infint  \u(\x,\t)\,\d{\x}.  \eqlabel{intdiff}
\end{eqnarray}
Function $\fit(\x)$ plays the role of the relative fitness of the
subpopulation with trait value $\x$.  Considering this function a
constant would not be appropriate as it would remove any selection; in
our simplified version we take it to be a linear function
$\fit(\x)=\fitz+\fitg\x$, where $\fitz,\fitg=\const$. This results in
the equation
\begin{equation}
  \df{\u}{\t} =  ( \fitz + \fitg\x - \N(\t) ) \u + \mut \ddf{\u}{\x}, 
  \quad
  \N(\t) = \infint  \u(\x,\t)  \,\d{\x} ,          \eqlabel{intdifflin}
\end{equation}
which is essentially identical e.g. to equation (2.1)
postulated by~\citet{Calsina-Perello-1995}.
Substitution 
\begin{equation}
  \u(\x,\t) = \N(\t) \, \pdf(\x,\t),                \eqlabel{normalize}
\end{equation}
brings the evolution equation~\eq{intdiff} to the well 
known~\citep[e.g.][]{Taylor-Jonker-1978,Page-Nowak-2002,Karev-2010}
``replicator-mutator'' form,
\begin{equation}
  \df{\pdf}{\t} = \left( \fit(\x)-\fitm(\t) \right) \pdf + \mut
  \ddf{\pdf}{\x}, 
  \qquad
  \fitm(\t) = \infint \pdf(\x,\t) \,\fit(\x) \,\d\x .   \eqlabel{repmut}
\end{equation}
Note that by virtue of~\eq{normalize} and the definition of $\N$
in~\eq{intdiff}, we automatically have
\begin{equation}
  \infint\pdf(\x,\t)\,\d\x=1                            \eqlabel{pdfunittotal}
\end{equation}
at any time. 
Function $\pdf(\x,\t)$ is the probability density function (PDF) of the
population in the trait space $\x$ at time $\t$, and $\fitm(\t)$ is
the mean fitness of the population at that time. If a solution of the
replicator-mutator equation~\eq{repmut} is known, then the total
population size can be found by solving equation
\begin{equation}
  \Df{\N}{\t} = ( \fitm(\t) - \N ) \N,                      \eqlabel{ODEsize}
\end{equation}
which then recovers a solution to the original evolution
equation~\eq{intdiff} via~\eq{normalize}.

A further substitution
\begin{equation}
  \pdf(\x,\t) = \P(\t) \, \v(\x,\t)                         \eqlabel{linearizing}
\end{equation}
brings equation~\eq{intdiff} to the form
\begin{equation}
    \df{\v}{\t} = \fit(\x)\,\v + \mut\,\ddf{\v}{\x},        \eqlabel{linpde} 
\end{equation}
provided that the factor $\P$ satisfies
\begin{equation}
  \dot\P/\P = -\P\,\infint\fit(\x)\,\v(\x,\t)\,\d{\x}.   
\end{equation}
Hence the integral part of the equation uncouples from the
differential, the closed linear differential equation~\eq{linpde} can be
solved first, and the linearizing factor can be found afterwards
via~\eq{linearizer} which, for the initial condition $\P(0)=1$, gives
the explicit expression
\begin{equation}
  \P(\t) = \left( 
    1 + \int\limits_0^{\t}\infint \v(\x,\tf) \,\fit(\x) \,\d\x\,\d\tf
  \right)^{-1}                                              \eqlabel{linearizer}
\end{equation}
(this is a continuous-trait variant of linearization used
by~\citet{Schuster-2011}). 
In ecological terms, the linear equation~\eq{linpde} corresponds to
the case when the subpopulations with different traits $\x$ are in no
direct competition with each other or even within themselves and,
aside from mutations described by the term $\mut\v_{\x\x}$, each
subpopulations multiplies by a Malthusian law with its own reproduction
rate $\fit(\x)$. 

For the linear fitness function $\fit(\x)$, we have the alternative
forms of~\eq{repmut} and \eq{linpde} respectively as
\begin{equation}
  \df{\pdf}{\t} =  ( \fitz + \fitg\x - \fitm(\t) ) \, \pdf + \mut \ddf{\pdf}{\x}, 
  \quad
  \fitm(\t) = \infint \fit(\x)\,\pdf(\x,\t) \,\d{\x} ,      \eqlabel{repmutlin}
\end{equation}
and
\begin{equation}
    \df{\v}{\t} = ( \fitz + \fitg\x)\,\v + \mut\,\ddf{\v}{\x} . \eqlabel{linpdelin} 
\end{equation}
In the next two sections, we concentrate on the solution of
equation~\eq{intdifflin} and its quivalent forms~\eq{repmutlin}
and~\eq{linpdelin}, before relaxing the simplifying assumptions in
\secn{perturbation}.

\section{The normal solution}
\secnlabel{normal}

We look for
solutions that describe the dynamic change of the 
population during its gradual adaptation, while far from any
evolutionary steady state.
We note that equation~\eq{repmutlin} admits a family of exact self-similar
solutions, which are Gaussians, or PDFs
of normal distributions,
\begin{equation}
   \pdf(\x,\t) = \PDF(\x,\t \,;\, \mean,\stddev)
   = \frac{1}{\stddev(\t)\sqrt{2\pi}} \, \exp\left[ -  \frac{
      (\x-\mean(\t))^2
    }{
      2\stddev(\t)^2
    }
  \right] .                                                 \eqlabel{normsol}
\end{equation}
As is easily verified by direct substitution, 
function \eq{normsol} is a solution of equation \eq{repmutlin}, provided that the parameters 
of the normal distribution obey the following system of ordinary
differential equation:
\begin{subequations}                                        \eqlabel{ODE}
  \begin{align}
    \Df{\mean}{\t} &= \fitg\,\stddev^2 ,                      \eqlabel{ODEmean}\\
    \Df{\stddev}{\t} &= \mut/\stddev  .                         \eqlabel{ODEdisp}     
  \end{align}
\end{subequations}
We shall call~\eq{normsol} a
\emph{normal} solution of~\eq{intdifflin}.

\myfigure{ \centerline{\includegraphics{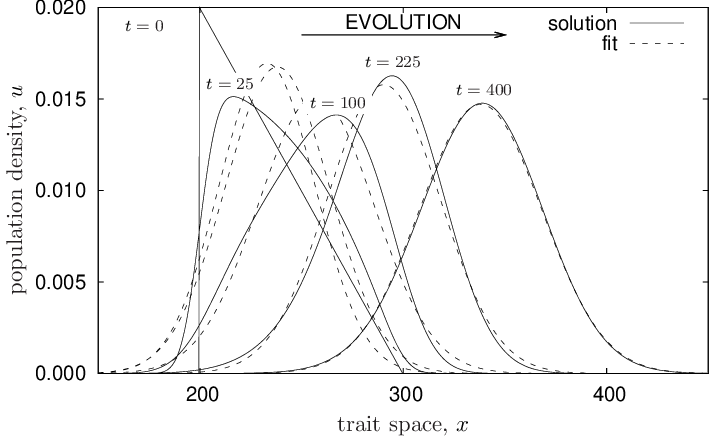}} }{%
  Establishment of the self-similar solution in \eq{intdifflin}%
}{%
  Profiles of population density in the trait space at selected
  moments of time during initial transient following a ``triangular''
  initial distribution.  The dashed lines are normal distributions
  corresponding to the solutions at chosen moments of time (with the
  same mean and variance).  Parameters: $\fitz=1$, $\fitg=1/3000$,
  $\mut=1$. Numerical simulation on the interval $\x\in[0,3000]$ with
  Neumann boundary conditions (simulation with Dirichlet boundary
  conditions or wider interval produces indistinguishable results),
  timestepping by forward Euler with second order accurate central
  difference for $\partial^2\u/\partial \x^2$ and trapezoidal rule for
  $\int \u \,\d{\x}$, space step $1/4$, time step $1/50$. Shown is
  only the left end of the interval of $\x$. %
}{Fig1}

We stress here that functional form~\eq{normsol} is not an arbitrary
assumption, but an exact consequence of the evolution
equation~\eq{intdifflin}, once appropriate initial conditions are
supplied.  These initial conditions should, of course, be Gaussian.
However, numerical simulations shown in~\fig{Fig1} suggest that the
general solution at arbitrary initial distributions asymptotically
become normal as time increases, so the special class~\eq{normsol}
should in fact be fully representative. This is indeed the case as
we show below. 

\section{General solution}
\secnlabel{general}

We formulate properties of the solution to~\eq{repmutlin} for
a wide class of initial conditions, which generalize the properties of the
normal solutions shown above. We do that in terms of the moments
of the function $\pdf(\x,\t)$ considered as the PDF
of the random quantity $\x$ at a given moment of time
$\t$. Namely, we use the mean, 
\begin{equation}
  \mean(\t) = \infint \x\,\pdf(\x,\t)\,\d\x,                \eqlabel{mean}
\end{equation}
and the variance
\begin{equation}
  \stddev^2(\t) = \infint (\x-\mean(\t))^2 \,\pdf(\x,\t)\,\d\x .
                                                            \eqlabel{variance}
\end{equation}

\begin{theorem}\label{thm-one}
  Let $\pdf(\x,\t)$ be a solution of~\eq{repmutlin} with initial
  condition $\pdf(\x,0)=\pdfini(\x)$. We assume that 
  that the initial
  condition $\pdfini(\x)$
  \begin{enumerate}[label=(A\arabic{*})]
  \item\label{Atotalone}    is normalized, $\infint\pdfini(\x)\,\d\x=1$, 
  \item\label{Asupport}     has a finite support, $\pdfini(\x)=0$ for $\x\not\in[\xm,\xp]$,
  \item\label{Aanalytical}  is analytical for $\x\in[\xm,\xp]$.
 \end{enumerate}
  Then this solution can be written in the form
  \begin{equation}
    \pdf(\x,\t) = \frac{1}{\stddev(\t)} \w\left(\frac{\x-\mean(\t)}{\stddev(\t)},\t\right)    
    \eqlabel{first}
  \end{equation}
  where $\w(\cdot,\t)$ is a PDF of a zero-mean, unit-variance distribution, and
  for all $\t\ge0$, 
  \begin{enumerate}[label=(C\arabic{*})]
    \item\label{C-distribution} $\w(\cdot,\t)$
    has moments of all orders, 
    \[ 
    \Moment_\n(\t)=\int \xs^\n \,\w(\xs,\t) \, \d\xs, 
    \qquad
    |\Moment_n(\t)|<\infty,
    \qquad
    \n=1,2,3,\dots,
    \]
  \item\label{C-convergence}
      it converges (in distribution) to the PDF of the normal distribution,
      \[
        \lim\limits_{\t\to\infty}\w(\xs,\t) 
        = \frac{1}{\sqrt{2\pi}} \e^{-\xs^2/2},                   
      \]
   \item\label{C-mean}
      the mean of PDF $\pdf(\cdot,\t)$ varies according to
      \begin{equation}
        \Df{\mean}{\t} = \fitg \stddev^2,         \eqlabel{dotmean}
      \end{equation}
    \item\label{C-var}
      and its variance according to
      \begin{equation}
        \Df{\stddev}{\t} = \frac{\mut}{\stddev} + \frac12\fitg\stddev^2\skew(\t),
                                                  \eqlabel{dotstddev}
     \end{equation}
      where $\skew(\t)\equiv\Moment_3(\t)$. 
    \end{enumerate}
\end{theorem}

The proof is given in \appx{AppB}. 

Conclusion~\ref{C-mean} states that equation~\eq{ODEmean} remains
exact in the general case. This is, of course, a special case of the
generic Fisher-Price law~\citep{Fisher-1930,Price-1972} as applied to
the current model.

On the contrary, conclusion~\ref{C-var} states that
equation~\eq{ODEdisp} is not exact and in the general case requires a
correction associated with the instant value of skewness $\skew$ of
the distribution.  However, \eq{ODEdisp} is ``asymptotically valid''
for large $\t$, as the skewness, according to~\ref{C-convergence}
vanishes in the limit~$\t\to\infty$. Moreover, from the proof we can
see that the asymptotic order of $\skew$ is such that the second term
in~\ref{C-var} is asymptotically smaller than the first term in the
limit~$\t\to\infty$.

\section{Perturbation theory}
\secnlabel{perturbation}

The previous results were for a simplified version of the model, where
dependencies $\C(\x)$, $\D(\x)$ were replaced by constants and
$\fit(\x)$ was replaced by a linear function, to allow a simple
analytical solution. Now we would like to ensure that these solutions
are not exceptional and small violation of the simplifying assumptions
will not lead to completely different solutions. So, we now consider
generic smooth dependencies for $\C(\x)$, $\D(\x)$ and $\fit(\x)$, but
assume that the variation of $\x$ across the population at any given
time is smaller than the typical scale at which these functions vary
significantly. So, we develop a perturbation theory where the small
parameter is $\stddev$, the standard deviation of the selected trait
in the population. Admittedly $\stddev$ is not a constant but a
dynamic variable; this however does not affect the formal asymptotic
expansions. Equivalently one could use the inital value $\stddevi$ as
the small parameter. This, however, complicates notation, so we do it
only in one place where such explicit treatment is essential.  We
require that functions $\C(\x)$, $\D(\x)$ and $\fit(\x)$ and the
necessary number of their derivatives are bounded and that $\D(\x)$ is
everywhere bigger than some positive constant, so that
$|\D'(\x)/\D(\x)|$ is bounded.  We silently assume that all the
moments are at most $\O{1}$, before establishing their actual
asymptotic orders more accurately. All results in this section are
obtained formally, without any attempts of rigorous justification.

We start from the integro-partial differential
equation~\eq{intdiffstart} 
and bring it to the form of replicator-mutator equation by
substitution
\begin{equation}
  \u(\x,\t) = \N(\t) \, \pdf(\x,\t)
\end{equation}
which gives
\begin{equation}
  \pdf_\t = \left( \fit(\x) - \fitm(\t) \right) \pdf 
  + \left[ \Ccoeff(\x)\pdf + \Dcoeff(\x)\pdf_\x \right]_\x,
  \qquad
  \fitm(\t) = \int \fit(\x) \pdf(\x,\t)\,\d{\x}. 
\end{equation}

Now we pass from the probability density $\pdf(\x,\t)$ to the
normalized (zero-mean, unit-variance) probability density $\w$, through
$\pdf$'s mean
\begin{equation}
  \mean(\t) = \int \x \pdf(\x,\t) \,\d{\x}, 
\end{equation}
and its variance
\begin{equation}
  \stddev^2(\t) = \int (\x-\mean(\t))^2 \pdf(\x,\t) \,\d{\x}, 
\end{equation}
via substitution
\begin{equation}
  \pdf(\x,\t) = \frac{1}{\stddev(\ts)} \, \w(\xs,\ts),
  \qquad
  \xs=\frac{\x-\mean(\t)}{\stddev(\t)}, 
  \quad
  \ts = \t.
\end{equation}
(for chain rule differerentiation, it is convenient to distinguish the
time variables in the old coordinates $(\x,\t)$ and the new
coordinates $(\xs,\ts)$; for functions of one variable such as
$\stddev$ and $\mean$ this distinction is of course not important).
Then
\begin{equation}
  \w_\ts = \frac{\dot\stddev}{\stddev}\,\w
  + \left( 
    \frac{\dot\mean}{\stddev} + \xs \frac{\dot\stddev}{\stddev} 
  \right) \w_\xs
  + \left( \ftilde - \fitm \right) \,\w
  + \Ftilde_\xs,                                  \eqlabel{w-eqn}
\end{equation}
where
\(\displaystyle
  \Ftilde(\xs,\ts)
  = \frac{1}{\stddev} \Ctilde \w
  + \frac{1}{\stddev^2} \Dtilde w_\xs
\), \(\displaystyle
 \Ctilde(\xs,\ts)=\Ccoeff\left(\mean(\ts)+\stddev\xs\right) 
\), \(\displaystyle
  \Dtilde(\xs,\ts)=\Dcoeff\left(\mean(\ts)+\stddev\xs\right) 
\), \(\displaystyle
 \ftilde(\xs,\ts)=\fit\left(\mean(\ts)+\xs \stddev\right) 
\), \(\displaystyle
 \fitm(\ts)=\int \ftilde(\xs,\ts) \w(\xs,\ts) \,\d\xs 
\).
By construction, equation \eq{w-eqn} is subject to
constraints
\begin{equation}
  \int \w(\xs,\ts) \,\d{\xs} = 1,                 \eqlabel{totconst}
\end{equation}
\begin{equation}
  \int \xs \, \w(\xs,\ts) \,\d{\xs} = 0 ,         \eqlabel{meanconst}
\end{equation}
\begin{equation}
  \int \xs^2 \, \w(\xs,\ts) \,\d{\xs} = 1 .       \eqlabel{varconst}
\end{equation}

Let us expand functions of $\x$ in Taylor series,
\[
  \ftilde(\xs,\ts) 
  = \sum\limits_{\n=0}^{\infty} \frac{1}{\n!} \stddev^{\n} \xs^{\n} \fit_{\n},
  \qquad \fit_{\n} = \fit_{\n}(\ts) \bydef \fit^{(\n)} (\mean(\ts)),
\]\[
  \Ctilde(\xs,\ts) 
  = \sum\limits_{\n=0}^{\infty} \frac{1}{\n!} \stddev^{\n} \xs^{\n} \C_{\n},
  \qquad \C_{\n} = \C_{\n}(\ts) \bydef \C^{(\n)}(\mean(\ts)),
\]\[
  \Dtilde(\xs,\ts) 
  = \sum\limits_{\n=0}^{\infty} \frac{1}{\n!} \stddev^{\n} \xs^{\n} \D_{\n},
  \qquad \D_{\n} = \D_{\n}(\ts) \bydef \D^{(\n)}(\mean(\ts)),
\]
and consider the formal asymptotic expansion of equation~\eq{w-eqn} in the
small parameter $\stddev$.

It is straightforward to see that if $\int \w\,\d{\xs}=1$ at $\ts=0$,
it remains so for all $\ts>0$, 
so constraint~\eq{totconst} is always
satisfied. 
The constraints~\eq{meanconst} and \eq{varconst} lead to asymptotic
series for $\dot\mean$ 
and $\dot\stddev$.  
Further, multiplying both
sides of equation~\eq{w-eqn} by $\xs^\n$, $\n\ge3$, and integrating over $\xs$
leads to asymptotic series for the moments $\Moment_{\n}$. 
In this
way, we obtain
\begin{align}
  \dot{\mean}
  &
  = ( \D_1-\C_0 ) 
  + \left( \fit_1 + \frac12\D_3-\frac12\C_2  \right) \stddev^2
  + \left( \frac12\fit_2 + \frac16\D_4-\frac16\C_3 \right) \skew\stddev^3 
  + \O{\stddev^4},
  \nonumber\\
  \dot\stddev
  &
  = 
    \D_0 \stddev^{-1} 
  + \left( \frac32 \D_2
  - \C_1 \right) \stddev
  + \left(
    \frac12 \fit_1
    - \frac12 \C_2 
    + \frac{2}{3} \D_3
  \right) \skew \stddev^2
  + \O{\stddev^3} ,
  \nonumber\\
  \dot\skew
  &
  =
  - 3 \D_0\skew \stddev^{-2} 
  + 6 \D_1 \stddev^{-1} 
  + \frac{3}{2} \D_2 \skew \stddev^{0}
  + \O{\stddev^1} .
  \eqlabel{as-three}
\end{align}
So, in the selected asymptotic orders, of all the moments only the skewness
$\skew=\Moment_3$ affects the evolution
speed. To see how big is its effect, we now need to consider
$\stddev=\stddev(\t)$ as a function of time rather than merely a small parameter. 
To estimate the upper bound for $\skew(t)$, we 
consider $\skew$ as a function of $\stddev$, with the initial
condition $\skew(0)=\skewi$, $\stddev(0)=\stddevi$:
\[
  \Df{\skew}{\stddev} = \dot\skew/\dot\stddev
  =
  - 3 \skew \stddev^{-1}  + 6 \D_1\D_0^{-1} 
    + \O{\stddev^1} ,
\]
hence
\[
  \stddev^3 \skew - \stddevi^3 \skewi
  = 6 \int\limits_{\stddevi}^{\stddev}
  \left(\D_1/\D_0\right)\stddev'^3\,\d\stddev' 
  + \O{\stddev^5} ,
\]
where the integrand $\D_1/\D_0$ depends on $\stddev$
via $\mean=\mean(\t)$ and $\t=\t(\stddev)$. 
Let us assume that 
\[
  |\D_1/\D_0|\le \gDmax = \const
\]
for the whole solution under consideration. Then
it follows that
\[
  |\skew| \le \left|\skewi (\stddevi/\stddev)^3\right| + \frac32\gDmax\stddev
  + \O{\stddev^2}.
\]
So, our upper bound for skewness $\skew$ consists of two components:
one is related to the decaying contribution of the initial condition,
$|\skewi (\stddevi/\stddev)^3|<|\skewi|$, and
$\skewi (\stddevi/\stddev)^3=\O{\t^{-3/2}}\to0$ as $\t\to\infty$, 
and the other is the contribution
of the perturbation, which is $\O{\stddev}$. So if the effect of the
initial skewness can be neglected, say $\skewi=\O{\stddev}$, then we have
$\skew=\O{\stddev}$, and from~\eq{as-three} 
we have finally our asymptotic evolution equations

\begin{subequations} \eqlabel{as-two}
\begin{align}
  \dot{\mean}
  &
  = ( \D'(\mean)-\C(\mean) ) 
  + \left( \fit'(\mean) + \frac12\D'''(\mean)-\frac12\C''(\mean)  \right) \stddev^2
  + \O{\stddev^4} ,
  \\
  \dot\stddev
  &
  = 
    \D(\mean) \stddev^{-1} 
  + \left( \frac32 \D''(\mean)
  - \C'(\mean) \right) \stddev
  + \O{\stddev^3} .
\end{align}
\end{subequations}
This (asymptotically) closed system of ordinary differential is a
generalization of the previous result~\eq{ODE} and it asserts that
that this simple system of two equations remains approximately valid even
if fitness gradient and mutation diffusivity are not constant, only
subject to the drift term due to mutation bias $-\C(\mean)$, as we
discussed above. Moreover, it provides higher-order corrections to
that simple system, if necessary.  

Further increase in asymptotic accuracy will require including into
consideration higher-order moments.  Let us consider higher-order
asymptotic equations for a special case when $\Dcoeff(\x)=\D=\const$,
$\Ccoeff(\x)=\C=\const$ so only the fitness $\fit(\x)$ is
trait-dependent. Then reasoning as before, we see that in this case
$\skew=\O{\stddev^3}$, 
$\Moment_4=3+\O{\stddev^3}$, and therefore
\begin{subequations} \eqlabel{as-two-other}
\begin{align}
  \dot\mean & = 
  -\C 
  + \fit'(\mean)\stddev^2
  + \frac12\skew\fit''(\mean)\stddev^3
  + \frac12\fit'''(\mean)\stddev^4
  + \O{\stddev^5} ,
  \\
  \dot\stddev &=
  \D\stddev^{-1}
  + \frac12\skew\fit'(\mean)\stddev^2
  + \frac12\fit''(\mean)\stddev^3
  + \O{\stddev^4} .
\end{align}
\end{subequations}
The third term in the right-hand side of~\eq{as-two-other}(b)
describes the effect on the variance of the stabilizing ($\fit''<0$)
or disruptive ($\fit''>0$) selection. Note that at a fitness maximum,
$\fit'=0$, $\fit''<0$, equation \eq{as-two-other}(b) gives a
stationary variance at
$\stddev^2=\left(-2\D/\fit''(\mean)\right)^{1/2}$, in agreement with
the classical result by \citet{Kimura-1965}.

\section{Discussion}
\secnlabel{discussion}

We have considered solutions of a simple model of gradual Darwinian evolution in
continuous phenotypicl trait space and continuous time, while far away
from evolutionary stable equilibrium. 

\myfigure{
  \centerline{\includegraphics{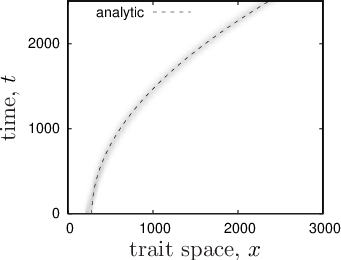}}
}{%
  Accelerating evolution according to equation \eq{intdifflin}.%
}{
  Continuation of the same simulation as shown in~\fig{Fig1}, 
  on a larger scale.
  The solution of~\eq{intdifflin} is shown as density plot on the
  space-time plane. 
  For comparison, the dashed line shows the
  corresponding solution $\x=\mean(\t)$ of equations~\eq{ODE}, with
  initial condition $\stddev(0)=0$ and $\mean(0)$ at the fittest edge
  of the initial condition of~\eq{intdifflin}. 
}{Fig2}

In biological terms, the main predictions of the model are:
\begin{itemize}
\item Equation~\eq{ODEsize} states that total population size $\N$ is
  described by Verhulst dynamics, where the mean reproduction rate
  $\fitm(\t)$ is itself dynamically changing.
\item Equations~\eq{ODEmean}, \eq{dotmean}, \eq{as-two}(a),
  \eq{as-two-other}(a) describe the dynamics of the mean trait value
  $\mean(\t)$, which has a selective component, proportional to the
  local fitness gradient $\fit'(\mean)$ and instant distribution
  variance $\stddev^2(\t)$, and a non-selective component, due to
  mutation bias $\Ccoeff(\mean)$. The selective component corresponds
  to the canonical equation~\eq{canonical} of the adaptive dynamics
  and to Fisher's Fundamental Theorem of Natural Selection. That is,
  not only population evolves faster in a stronger fitness gradient,
  but also populations which are more diverse with respect to the
  selected trait, evolve faster, and those that are very homogeneous
  in the selected trait, evolve slower.
\item Equations~\eq{ODEdisp}, \eq{dotstddev}, \eq{as-two}(b), \eq{as-two-other}(b) describe
  the dynamics of the variance. It describes diffusive spread of the
  variance, in the long run by the law $\stddev^2=2\mut\t$, where the mutations rate
  $\mut$ plays the role of diffusivity. 
\item The Gaussian distribution of the trait in the population emerges
  spontaneously during the course of evolution, as stated by
  conclusion~\ref{C-convergence} of Theorem~\ref{thm-one}.  This
  eliminates need for any artificial closing procedures in the moments
  equations.  The Gaussian distribution is preserved even if the
  gradient of fitness and the mutation parameters vary with the change
  of the trait coordinate, provided that their variation is slight
  within the spread of the population.
\end{itemize}

Since the rate of evolution is proportional to the variance, and the
variance in the long run is constantly growing, these results predict that the
evolution has a propensity to accelerate, for as long as the
assumptions of the model remain valid. This property is illustrated by
a simulation presented in \fig{Fig2}.

In biological terms, the growth of the variance
  $\stddev^2(\t)$ in the course of evolution is due to absence of
  stabilizing selection in the simplified version of our model. 
  Stabilizing selection can stop that growth, as illustrated by the
  stationary solution of equations \eq{as-two-other}
  around a local optimum in the
  fitness landscape. In that example, however, the drift of the
  mean $\mean(\t)$ also stops.  \citet{Hallatschek-2011}
  describes stationary propagating wave solutions which are
  characterized, in our notations, by a stationary $\stddev^2$ but a 
  constantly changing $\mean$. He
  find such solutions in a class of evolution equations broadly
  similar to~\eq{intdiffstart} but with both the environmental
  selection and the intraspecific competition changing in time,
  synchronously with the progress of the wave.  In the evolutionary
  context, such synchronous change may be provided during
  co-evolution, when change of phenotypical distribution of one
  population contributes to changes of the fitness landscapes of other
  populations.

Our underlying assumptions are self-limiting. If coefficients
$\Ccoeff$, $\Dcoeff$ and $\fit'$ vary significantly over the
trait space scale of $\xc$, then our perturbation theory will remain
valid only for as long as the variance is small enough,
$\stddev^2\ll\xc^2$.  On the other hand, this variance grows
unbounded, $\stddev^2\approx2\mut\t$, hence this theory may be only
valid for a limited time interval, $\t\ll\xc^2/(2\mut)$.  However,
this limitation is consistent with our goal which is to consider the
transient phase of the evolution, before the population reached the
optimum, as stationary distribution around the optimum is considered
in other works.

To conclude, we discuss the role of mutations, specifically the value
of coefficient $\Dcoeff$ in the above results. According to
equation~\eq{dotmean}, the instantaneous rate of change of the average
value of the trait does not depend directly on $\Dcoeff$, which is
entirely consistent with the general Fisher-Price law which claims
that it is determined only by intensity of selection and variance of
the instant trait distribution. As to the variance itself, then
according to \eq{dotstddev} in the long run it always grows with the
rate directly proportional to $\Dcoeff$. The convergence of the
distribution shape to Gaussian also depends on mutation. According to
estimates obtained in \appx{AppB},
$\skew(t)\approx \Qconst/(\fitg^3\mut^{3/2}\t^{9/2})$, $t\to\infty$,
where $\Qconst$ is a constant depending on initial conditions. Then,
accepting smallness of skewness $\skew$ as a measure of convergence to
Gaussian distribution, we see that this convergence is due to both
selection ($\fitg$) and mutation ($\mut$) and so will be very slow in
the limit of $\fitg\to0$ or $\mut\to0$. So, for a fixed time interval
and very near to the fitness extremum, or for very small mutation
rate, our results become inapplicable. This is of little surprise as
mutation without selection as well as selection without mutation are
completely different biological situations as well as completely
different mathematical problems.  The case of small mutation rate is
considered in a recent paper by \citet{Sasaki-Dieckmann-2011}, who
start from selection without mutation and then add mutation as a
perturbation, thus presenting an approach which is complementary to
the one given here, and more appropriate for analysis near fitness
extrema, as opposed to nearly-linear fitness landscapes considered
here.

\bibliographystyle{spbasic}      

\newpage
\appendix
\section{Population dynamics origin for the model}
\appxlabel{AppA}

Many if not all individual ideas of our derivation are found in
literature; however we could not find such derivation as a whole, so
present it here in its entirety, step by step, emphasizing all
assumptions made on the way, in order to identify the limitations of
the resulting model, which may lead to ways to overcome these
limitations.  We start from the population dynamics model of
Lotka-Volterra-Gause type, where the predator population consists of a
number of subpopulations $\zoo_\j$, differing in their parameters, all
consuming the same prey (resource) $\Prey$, mutating into each other,
and not interacting otherwise:
\begin{eqnarray}
  \Df{\Prey}{\t} &=& \Prey \left( 
    \alp - \sum\limits_{\j=-\infty}^{\infty} \bet_\j \zoo_\j  - \Prey
  \right) , \nonumber \\
  \Df{\zoo_\j}{\t} &=& \zoo_\j \left( 
    - \gam_\j + \del_\j \Prey 
  \right) + \Mut_\j,
  \qquad \j=0\pm1,\pm2,\dots ,
           \eqlabel{PZ}
\end{eqnarray}
where $\alp$ is the low-density reproduction rate of the resource in
absence of grazing pressure, $\bet_\j$ describes the per capita
grazing pressure by trait $\j$ on the common resource, the coefficient
at the quadratic term in the first equation of~\eq{PZ} is unity due to
the choice of the unit of measurement for $\Prey$,
$(-\gam_\j+\del_\j\Prey)$ is the resource-dependent reproduction rate
of trait $\j$, and $\Mut_\j$ is the contribution of mutations, defined
as
\begin{equation}
  \Mut_\j = 
  \Mut^+_{\j-1}\zoo_{\j-1}+\Mut^-_{\j+1}\zoo_{\j+1}
  - \left( \Mut^+_\j+\Mut^-_\j \right) \zoo_\j .  \eqlabel{mutations}
\end{equation}
Here we chose index $\j$ so it enumerates subpopulations monotonically
with respect to the selected trait.  The coefficients $\Mut^\pm_\j$
are the rates of mutation of subpopulation $\j$ that respectively
increase or decrease the trait index $\j$.  So, in this model
different types differ both by internal dynamics and by their
interaction with the common resource, and their competition is
indirect, only via dependence on the common resource.
The population model with mutations \eqtwo(PZ,mutations) is admittedly
very simple and ignores many biological aspects which can be very
important in real life.  For instance, it ignores frequency dependent
selection, as the population dynamics of each type does not depend on
its abundance relative to other types, but only on the ``mean field'',
represented by the common resource. Also, we shall consider ``smooth''
dependence of the coefficients in this model on the one-dimensional
index $\j$, which will effectively impose one-dimensionality on the
resulting model, which is also a significant restricting assumption,
see~\cite{Metz-etal-2008}.  We stress however that our purpose here is
not to derive a biologically realistic model, but only to present a
simple example of how equations of evolution can be derived from
population dynamics equations.

For mathematical simplicity, we take the set of subpopulations
$\{\j\}$ to be infinite, which in practice only means that the overall
diversity of the population observed during the stage of evolution
under consideration is much smaller than all that are theoretically
possible. Also for mathematical simplicity only, we take that
mutations are so small they occur only between subpopulations adjacent
in the trait index $\j$; this assumption is not essential: admitting
mutations to any \emph{small} distance in $\j$ leads to the same form
of dynamic equation in the continuous limit. What is important is that
we neglect the probability of mutations that change the trait
significantly, in comparison with the diversity of the population at
any given moment of time; taking those into account would make the
resulting equation a bit more complicated (mutations will be described
by an integral rather than differential term, see
e.g.~\cite{Champagnat-2006}).  There are of course also lethal
mutations but they can be considered as contributing to the death rate
and thus implicitly included in $\gam_\j$.

Our next assumption is that the dynamics of resource (prey) are much
faster than those of predators, so the resource concentration can be
adiabatically eliminated.  We find the quasi-stationary resource
concentration as
\[
  \Prey = \alp - \sum\limits_{\j=-\infty}^{\infty} \bet_\j \zoo_\j ,
\]
which gives the predators dynamics equations as
\begin{equation}
  \Df{\zoo_\j}{\t} = \zoo_\j \left( 
    \alp\del_\j - \gam_\j - \del_\j \sum\limits_{\j=-\infty}^{\infty}\bet_\j \zoo_\j 
  \right) + \Mut_\j.
  \eqlabel{discrete}
\end{equation}
We change the dynamic variables to measure the subpopulation sizes by
their grazing pressure, $\u_\j=\bet_\j\zoo_\j$, and turn to the
continuous limit via $\x=\j\h$, $\h\to0$, where $\x$ is a physical
measurement of the trait.  This leads to the integro-differential
equation
\begin{equation}
  \df{\u}{\t} = \left( 
    \hat\fit(\x) - \del(\x) \N(\t)
  \right)  \u 
  + \hat\Ccoeff(\x) \df{\u}{\x}
  + \hat\Dcoeff(\x) \ddf{\u}{\x} ,
  \qquad
  \N(\t)  =\infint  \u(\x,\t)\,\d{\x},            \eqlabel{intdiffhat}
\end{equation}
where 
\[
  \u(\x) = \lim \left(h \u_{\j}\right) ,
  \qquad
  \del(\x) = \lim \del_{\j} ,
\]\[
  \hat\fit(\x) = \lim\left(
    \alp\del_\j 
    - \gam_\j 
    + \Mut^+_{\j-1}\bet_\j/\bet_{\j-1}
    +\Mut^-_{\j+1}\bet_\j/\bet_{\j+1}
    - \Mut^+_\j -\Mut^-_\j 
 \right) ,
\]\[
  \hat\Ccoeff(\x) = \lim \h \bet_\j \left(
    - \Mut^+_{\j-1}/\bet_{\j-1}
    + \Mut^-_{\j+1}/\bet_{\j+1}
  \right) ,
\]\[
  \hat\Dcoeff(\x) =  \lim \frac12\h^2 \bet_\j \left(
    \Mut^+_{\j-1}/\bet_{\j-1}
    + \Mut^-_{\j+1}/\bet_{\j+1}
  \right) ,
\]
and all limits are taken as $\h\to0$, $\j=\x/\h$, $\x=\const$.  In
particular, we assume that dependence of mutation rate
$\Mut^{\pm}_{\j}$ on $\j$ and the asymmetry between beneficial and
deleterious mutation rates $\Mut^+_{\j}-\Mut^-_{\j}$ are such that
both limits $\hat\Ccoeff(\x)$ and $\hat\Dcoeff(\x)$ exist and
$\hat\Dcoeff(\x)\ne0$.  A change of variables
\[
  \Dcoeff(\x)=\hat\Dcoeff(\x), 
  \quad  
  \Ccoeff(\x)=\hat\Ccoeff(\x) - \Dcoeff'(\x),
  \quad  
  \fit(\x)=\hat\fit(\x) - \Ccoeff'(\x),
\]
transforms equation~\eq{intdiffhat} to 
\begin{equation}
  \df{\u}{\t} = \left( 
    \fit(\x) - \del(\x) \N(\t)
  \right)  \u 
  + \df{}{\x} \left[
    \Ccoeff(\x) \u
    + \Dcoeff(\x) \df{\u}{\x}
  \right],
  \qquad
  \N(\t)  =\infint  \u(\x,\t)\,\d{\x}. 
\end{equation}
Finally, in this paper, we consider the case $\del(\x)=\const$, and we
rescale population density so this constant is unity, which gives the
model~\eq{intdiffstart}.
In biological terms, this means that we assume that the competition
between different types is in the speed of reproduction which is
assumed strictly proportional to the carrying capacity, i.e. the
equilibrium population density of trait $\x$ in the given habitat in
absence of other types.  This choice is arbitrary and is made from
consideration of mathematical simplicity rather than motivated by any
specific biological examples.
We note however, that the effect of this, as well as many other
simplifying assumptions made above, can be relaxed by perturbation
theory, of the kind considered in \secn{perturbation}.

\section{Proof of Theorem~\ref{thm-one}}
\appxlabel{AppB}

First two brief preliminary remarks. 
\begin{itemize}
\item 
We shall use the identity~\eq{pdfunittotal}
in our proof. As already noted, this is ensured by construction of the
equation, and it also can be easily verified that once this property
is satisfied for the initial condition according to~\ref{Atotalone},
it is then preserved by the equation~\eq{repmut}.

\item
We shall, without loss of generality, assume
that $\xp=0$; otherwise, we just make transformation
$\x\mapsto\x-\xp$. 
\end{itemize}

Note also that some of the variables used here coincide by name with
variables used in \appx{AppA} but have different meaning
($\Mut$, $\dirac$). 

\begin{proof}

  $1^{\circ}$ We start by proving~\ref{C-distribution}.  We do that
  using the equivalent linear equation~\eq{linpdelin}. First, we
  construct the fundamental solution of that equation, that is, a
  generalized solution $\V(\x,\t\,;\,\xf)$ with initial condition
  $\V(\x,0\,;\,\xf)=\dirac(\x-\xf)$, where $\dirac()$ is Dirac's
  delta-function. This can be obtained from the normal solution
  $\PDF(\x,\t \,;\, \mean,\stddev)$~\eqtwo(normsol,ODE) with initial
  conditions $\mean(0)=\xf$ and $\stddev(0)=0$, transformed
  by~\eq{linearizing} and \eq{linearizer}, which leads to
  \begin{equation}
    \V(\x,\t\,;\,\xf) = \frac{1}{2\sqrt{\pi\mut\t}} \,\exp\left[
        (\fitz+\fitg\xf)\t + \frac{\mu\fitg^2\t^3}{3} - \frac{
          \left(\x-\xf-\fitg\mut\t^2\right)^2
        }{
          4\mut\t
        }
      \right].
  \end{equation}
  Hence for the generic initial condition $\v(\x,0)=\vini(\x)$ we have
  \begin{eqnarray}
    \v(\x,\t) &=& \infint \V(\x,\t\,;\,\xf) \, \vini(\xf) \, \d\xf 
    \nonumber\\
    &=& \Kone \infint
    \exp\left[ \frac{
        4\mbar\xf - (\x-\mbar-\xf)^2
      }{
        2\sbar^2
      }
    \right] \vini(\xf) \,\d\xf,                             \eqlabel{v-sol}
  \end{eqnarray}
  where 
  \[
  \Kone=\frac{
    \e^{\fitz\t+\mut\fitg^2\t^3/3}
  }{
    2\sqrt{\pi\mut\t}
  },
  \qquad
  \sbar^2 = 2\mut\t,
  \qquad
  \mbar = \fitg\mut\t^2
  \]
  are some known functions of time. Normalization of this solution
  gives the PDF $\pdf(\cdot,\t)$, the moments of which 
  are found as
  \[
  \Momv_\n=\Intv_\n/\Intv_0, 
  \qquad
  \Intv_\n=\infint \x^\n \infint
  \exp\left[ \frac{
      4\mbar\xf - (\x-\mbar-\xf)^2
    }{
      2\sbar^2
    }
  \right] \vini(\xf) \,\d\xf
  \,\d\x .
  \]
  The corresponding double integrals are absolutely convergent under
  the assumed properties of $\vini$. Furthermore, $\Intv_0>0$. Hence the existence
  of moments of all orders at all $\t$ for the PDF $\pdf$
  follows. In particular, the mean $\mean(\t)=\Momv_1(\t)$ and
  variance $\stddev^2(\t)=\Momv_2(\t)$ are defined for all $\t\ge0$.
  The normalized PDF $\w(\xs,\t)$ is obtained from
  $\pdf(\x,\t)$ via substitution $\x=\mean+\stddev\xs$, and 
  is zero-mean
  and unit-variance by construction. Its moments of all orders
  exist as they are the standardized moments of PDF $\pdf$. This
  is the conclusion~\ref{C-distribution} of the Theorem.

  $2^{\circ}$
  Now that the existence of $\mean(\t)$ and $\stddev(\t)$ for
  all $\t\ge0$ is established, we can proceed to prove the
  predictions \ref{C-mean} and \ref{C-var} about their dynamics. 
  To find the rate of change of the mean $\mean(\t)$, let us first note
  that for a linear fitness function $\fit(\x)$, we have
  \begin{equation}
  \fitm(\t) = \infint (\fitz + \fitg\x) \,\pdf(\x,\t) \,\d\x
  = \fitz + \fitg\mean(\t).                                     \eqlabel{fitmean}
  \end{equation}
  Now let us multiply both sides of~\eq{repmutlin} by $\x$ and
  integrate them over $\x\in\Real$. This gives
  \[ 
  \Df{\mean}{\t} = \infint \x \df{\pdf}{\t} \,\d\x
  = \infint \left( \fitz x + \fitg \x^2 - \fitm \x \right) \, \pdf \,\d\x
  + \mut \infint \x  \ddf{\pdf}{\x} \,\d\x 
  \]\[
  = \fitz\mean - \fitm\mean + \fitg \infint \x^2 \, \pdf \,\d\x 
  \]
  where we used the definition of the mean~\eq{mean} and the integral
  proprotional to $\mut$ vanishes when integrated by parts (the limit
  $\lim_{x\to\pm\infty}\left[\x\@_x\pdf(\x,\t)\right]=0$ can be verified
  using~\eq{v-sol}).
  Now we do an equivalent transformations in this equation,
  \[
  \Df{\mean}{\t} = \fitz\mean-\fitm\mean
  + \fitg\infint\left( \x^2 - 2\x\mean + \mean^2\right) \,\pdf\,\d\x
  + 2\fitg\mean\infint\x\,\pdf\,\d\x-\fitg\mean^2\infint\pdf\,\d\x
  \]\[
  = \fitz\mean-\fitm\mean +  \fitg\infint(\x-\mean)^2\,\pdf\,\d\x + 2\fitg\mean^2-\fitg\mean^2
  = \fitg\stddev^2 + (\fitz\mean+\fitg\mean^2) - \fitm\mean
  \]
  where we have used definitions~\eq{pdfunittotal}, \eq{mean} and \eq{variance}. 
  According to~\eq{fitmean}, the last two terms cancel out, which
  delivers the conclusion~\ref{C-mean} of the theorem. 

  Transformation~\eq{first} ensures some (already mentioned) identities for
  $\w(\xs,\t)$, which we will now need stated explicitly. Namely, 
  \begin{equation}
    \infint\w(\xs,\t)\,\d\xs=1                              \eqlabel{unittotal}   
  \end{equation}
  follows from~\eq{pdfunittotal}, 
  \begin{equation}
    \infint\xs\,\w(\xs,\t)\,\d\xs=0                            \eqlabel{zeromean}
  \end{equation}
  follows from~\eq{mean} and 
  \begin{equation}
    \infint\xs^2\,\w(\xs,\t)\,\d\xs=1                         \eqlabel{unitvariance}
  \end{equation}
  follows from~\eq{variance}. 
  Substitution of~\eq{first}
  into the linear evolution equation~\eq{repmutlin}, with account of~\eqset(unittotal-unitvariance)
  and the already proved identity~\ref{C-mean},
  leads to the following differential equation
  \begin{equation}
    \df{\w}{\t} = \fitg\stddev\, \left(\xs\w + \df{\w}{\xs}\right)
    + \frac{\dot\stddev}{\stddev} \, \left(\w + \xs\df{\w}{\xs}\right)
    + \frac{\mut}{\stddev^2} \ddf{\w}{\xs}.                   \eqlabel{weq}
  \end{equation}
  By considering the first three moments
  of both sides of this equation, we see that the subspace of
  functions defined by~\eqset(unittotal-unitvariance) is an invariant
  set of this equation if and only if
  \begin{equation}
    \dot\stddev = \frac{\mut}{\stddev} + \frac12 \fitg\stddev^2\infint\xs^3\w(\xs,\t)\,\d\xs
              = \frac{\mut}{\stddev} + \frac12 \fitg\stddev^2\skew,
  \end{equation}
  which is conclusion~\ref{C-var} of the Theorem. 

  $3^{\circ}$
  It remains to prove~\ref{C-convergence}. 
  We do it by the method of moments.
  An exact expression for the
  normalized PDF $\w(\xs,\t)$ is obtained from \eq{v-sol}
  via
  substitution $\x=\mean+\stddev\xs$, where $\mean=\mean(\t)$ and $\stddev=\stddev(\t)$ are the
  true mean and standard deviation of the $\v$-distribution
  (as opposed to the ``rough guess'' values of the same, $\mbar(\t)$ and $\sbar(\t)$),
  and an appropriate normalization. This
  gives
  \[
  \w(\xs,\t) = \Ktwo \infint \exp\left[ \frac{
      4\mbar\xf - (\stddev\xs+\mean-\mbar-\xf)^2
    }{
      2\sbar^2
    }
  \right] \vini(\xf) \,\d\xf ,
  \]
  where $\Ktwo$ is a coefficient depending only on $\t$ but
  not on $\xs$, chosen so that $\infint\w\,\d\xs=1$. 

  The moments of PDF $\w$ are found as
  \[  \Moment_\n = \Integral_\n/\Integral_0, \]
  where
  \begin{equation}
  \Integral_\n = \frac{1}{\sqrt{2\pi}} \infint \xs^\n \left\{ \infint \exp\left[ \frac{
        4\mbar\xf - (\stddev\xs+\mean-\mbar-\xf)^2
      }{
        2\sbar^2
      }
    \right] \vini(\xf) \,\d\xf \right\}\,\d\xs.             \eqlabel{In}
  \end{equation}
  The already mentioned absolute convergence of double integrals \eq{In} means that 
  Fubini's theorem applies and we can
  change the order of integration. On doing so, and also introducing
  notations
  \[ 
  \Shift=(\xf+\mbar-\mean)/\sbar,
  \qquad 
  \xs=\frac{\sbar}{\stddev}(\z+\Shift)
  \]
  we get
  \[
  \Integral_\n = \left(\frac{\sbar}{\stddev}\right)^{\n+1}
  \infint \frac{\e^{-\z^2/2}}{\sqrt{2\pi}} \left[
    \infint  (\z+\Shift)^\n \vb(\xf,\t) \,\d\xf
    \right]\,\d\z,
  \]
  where $\vb$ is the initial PDF, ``biased'' by selection
  towards fitter trait values: 
  \[
  \vb(\xf,\t) = \vini(\xf) \e^{\fitg\t\xf}.
  \]
  For brevity, we shall now omit dependence on time, until we start
  studying the $\t\to\infty$ asymptotics. 
  By using the binomial formula, the moment integrals are rewritten as
  \[
  \Integral_\n  = \left(\frac{\sbar}{\stddev}\right)^{\n+1} 
  \sum\limits_{\k=0}^{\n} \Binom{\n}{\k} \A_{\n-\k} \B_{\k},
  \]
  where
  \begin{equation}
  \A_\m = \infint \frac{\e^{-\z^2/2}}{\sqrt{2\pi}}\, \z^\m \,\d\z = \left\{
    \begin{array}{ll}
      0,       & \textrm{if $\m$ is odd,} \\
      (\m-1)!!, & \textrm{if $\m$ is even}
    \end{array}
  \right.                                                   \eqlabel{normal-moments}
  \end{equation}
  are the moments of the standard normal distribution, and
  \[
  \B_\k = \infint \Shift^\k \, \vb(\xf)\,\d\xf
  = \sbar^{-\k} \sum\limits_{\l=0}^{\k} \Binom{\k}{\l} (\mbar-\mean)^{\k-\l}\,\VB_\l,
  \]
  where, in turn, $\VB_\l$ are the moments of the biased initial condition,
  \begin{equation}
  \VB_\l = \infint \xf^\l \, \vb(\xf) \,\d\xf
  = \infint \xf^\l \,\e^{\fitg\t\xf}\, \vini(\xf)\,\d\xf.    \eqlabel{biasedmoms}
  \end{equation}
  Combining these together, we obtain
  \[
  \Integral_\n = 
  \left(\frac{\sbar}{\stddev}\right)^{\n+1}
  \sum\limits_{\k=0}^{[\n/2]} \frac{\n!}{(2\k)!!(\n-2\k)!} \left(\frac{\mbar-\mean}{\sbar}\right)^{\n-2\k}
  \sum\limits_{\l=0}^{\n-2\k} \Binom{\n-2\k}{\l} (\mbar-\mean)^{-\l} \, \VB_\l ,
  \]
  where $[x]$ denotes the integer part of $x$. In particular,
  \[
  \Integral_0 = \frac{\sbar}{\stddev}\,\VB_0,
  \]
  hence for the moments we have
  \[
  \Moment_\n = 
  \left(\frac{\sbar}{\stddev}\right)^{\n}
  \sum\limits_{\k=0}^{[\n/2]} 
    \frac{\n!}{(2\k)!!(\n-2\k)!} \left(\frac{\mbar-\mean}{\sbar}\right)^{\n-2\k}
  \sum\limits_{\l=0}^{\n-2\k} 
    \Binom{\n-2\k}{\l} (\mbar-\mean)^{-\l} \, \frac{\VB_\l}{\VB_0}. 
  \]
  These expressions can be used for determining the true mean and
  variance in terms of the initial PDF, via the
  identities~\eq{zeromean} and~\eq{unitvariance}. Namely, we have
  \[
  \Moment_1 = \frac{1}{\stddev\VB_0}
  \left[ (\mbar-\mean)\VB_0 + \VB_1 \right]
  = 0, 
  \]
  hence
  \begin{equation}
    \mean = \mbar + \VB_1/\VB_0.                            \eqlabel{mean-mbar}
  \end{equation}
  Then, 
  \[
  \Moment_2 = \frac{1}{\stddev^2\VB_0}
  \left[ \VB_2 + \sbar^2\VB_0 - \VB_1^2/\VB_0 \right]
  =1,
  \]
  hence
  \begin{equation}
  \stddev^2 = \sbar^2 + \VB_2/\VB_0 - \VB_1^2/\VB_0^2.      \eqlabel{stddev-sbar}
  \end{equation}
  With account of these, finally we have an exact formula for the
  moments of $\w$ in terms of the initial PDF $\vini$: 
  \begin{equation}
    \Moment_\n = \left(
      \sbar^2+\frac{\VB_2}{\VB_0} - \frac{\VB_1^2}{\VB_0^2}
    \right)^{-\n/2} 
    \sum\limits_{(\l,\k):2\k+\l\le \n}
    \frac{\n!\;\sbar^{2\k}}{(2\k)!!\;\l!\; (\n-2\k-\l)!}
    \left(-\frac{\VB_1}{\VB_0}\right)^{\n-2\k-\l}
    \left(\frac{\VB_\l}{\VB_0}\right).                      \eqlabel{Moments}
  \end{equation}

  So, the problem of the $\t\to\infty$ asymptotics of the
  moments is reduced to asymptotics of $\VB_\n$. 
  In this limit, we have
  $\sbar=(2\mut\t)^{1/2}\to\infty$ and $\mbar=\fitg\mut\t^2\to\infty$. 
  We also introduce $\s=\fitg\t$ for brevity, and $\s\to\infty$,
  too. 
  
  In terms of $\s$, integrals $\VB_\n$ defined by~\eq{biasedmoms}
  are, up to the signs, the bilateral Laplace image of the initial
  distribution $\vini$ for $\n=0$, and derivatives of that image for
  $\n>0$. Asymptotics of Laplace images are known to be very sensitive to
  analytical properties of the originals. So at this point
  our assumptions~\ref{Asupport} and \ref{Aanalytical} become
  essential. In accordance with the second preliminary remark,
  we set $\xp=0$ and $\xm=-\W$, $\W>0$, without loss of generality. 
  Then the initial PDF has the form:
  \begin{equation}
    \vini(\x) = \left\{\begin{array}{ll}
        \sum\limits_{\m=0}^{\infty} \a_\m \, \x^\m, & \x\in[-\W,0], \\
        0, & \x\not\in[-\W,0].
      \end{array}\right.                                        \eqlabel{initanal}
  \end{equation}
  Then for the biased moments we have 
  asymptotic series
  \begin{equation}
    \begin{split}
      \VB_\l & 
      = 
      \sum\limits_{\m=0}^{\infty} \a_\m
      (-1)^{\m+\l}
      \frac{(\m+\l)!}{\s^{\m+\l+1}}
      + \est 
    \end{split}                                             \eqlabel{VBas}
  \end{equation}
  where $\est$ stands for ``exponentially small terms'', that is terms
  $\O{\e^{-\expdec\s}}$ for any $\expdec\in(0,\W)$. 
  
  Let $\q\ge0$ be the smallest power in series~\eq{initanal}, i.e. 
  $\a_\m=0$ for $\m<\q$ and $\a_\q\ne0$. Then from~\eq{VBas} we have
  \[
  \VB_\l=(-1)^{\l+\q}\a_\q(\l+\q)!\s^{-\l-\q-1}\left(1+\O{\s^{-1}}\right).
  \]
  In particular, 
  \[
  \VB_1/\VB_0=-(\q+1) \s^{-1}\left(1+\O{\s^{-1}}\right)=\O{\t^{-1}}
  \]
  so \eq{mean-mbar} gives
  \begin{equation}
    \mean = \mbar+\O{\t^{-1}},                       \eqlabel{mean-mbar-as}
  \end{equation}
  and further
  \[
  \VB_2/\VB_0=(\q+2)(\q+1)\s^{-2}\left(1+\O{\s^{-1}}\right)=\O{\t^{-2}}
  \]
  so~\eq{stddev-sbar} gives
  \begin{equation}
    \stddev^2 = \sbar^2 + \O{\t^{-2}}.               \eqlabel{stddev-sbar-as}
  \end{equation}
  That is, the ``crude guesses'' do in fact give asymptotically
  correct predictions of the true mean and true variance. 
  This is only because we have chosen $\xp=0$, more about it later. 
 
  For $\l\ge2$ we have 
  \[
  \VB_\l/\VB_0 = 
  (-1)^{\l}
  \frac{(\q+\l)!}{\q!}
  \s^{-\l}
  \left( 1 + \O{\s^{-1}} \right).
 \]
 Substituting these results into \eq{Moments} we get, after some transformations,
  \begin{align}
    \Moment_\n =
    \left( 1 + \O{\s^{-1}} \right)
    (-1)^\n\n!
    \sum\limits_{\k=0}^{[\n/2]}
    \frac{1}{(2\k)!!\;\q!}
    \left(\frac{\q+1}{\sbar\s}\right)^{\n-2\k}\;
    \E_{\n-2\k},
  \end{align}
  where
  \begin{align}
    \E_\m = \sum\limits_{\l=0}^{\m}
    \frac{(\q+\l)!}{\l!\; (\m-\l)!} \left(\frac{-1}{\q+1}\right)^\l .
  \end{align}

 Here in the limit $\t\to\infty$, the dominant terms are those with the
 largest $\k$ such that $\E_{\n-2\k}\ne0$, and all others will be
 subsumed by the factor $\left(1 + \O{\s^{-1}}\right)$. 
 For odd $\n$, we have
   $\E_1=0$, $\E_3=-\frac13 \q!(\q+1)^{-2}$, so
 \begin{align} 
   \Moment_\n = 
    \frac{ \n!!(\n-1)(\q+1) }{ 3(\sbar\s)^3  }
    \left( 1 + \O{\s^{-1}} \right) 
    =\O{\t^{-9/2}}.                                            \eqlabel{Modd}
 \end{align}
 For even $\n$, we have $\E_0=\q!$
 which gives
 \begin{align} 
   \Moment_\n = & (\n-1)!! \, \left( 1 + \O{\t^{-1}} \right).  \eqlabel{Meven}
 \end{align}
 Summarising, we have for all $\n$ that
 \[
 \Moment_\n = \A_n + \O{\t^{-1}}, \qquad \t\to\infty, 
 \]
 where $\A_n$ are the moments of the standard normal distribution, \eq{normal-moments}.
 Since the normal distribution
 is uniquely characterized by its moments, convergence of $\w$ in moments here
 implies convergence of distributions, and we have the
 claimed~\ref{C-convergence}.
 \qed
\end{proof}

Two final remarks. 

\begin{itemize}
\item 
  Equations~\eq{mean-mbar-as} and \eq{stddev-sbar-as}, meaning that
  the ``crude guesses'' $\mbar$ and $\sbar$ happen to be
  asymptotically accurate for the true $\mean$ and $\stddev$, are a
  direct consequence of the choice $\xp=0$, as can be verified by
  repeating the calculations with generic $\xp$. There is a simple
  interpretation of this fact in biological terms. In the initial PDF
  $\pdfini(\x)$ there is only a finite range $\x\in[\xm,\xp]$ of
  traits present, and the descendants of individuals with different
  traits make different contributions to the overall population at
  different times; but in the present model, the distribution of
  traits in the population in the long run is such as if descendants
  of the fittest ancestors, $\x=\xp$, dominated in it, regardless of the
  effect of mutations. Mathematically, this is eventually down to
  linearity of~\eq{linpde}. With initial conditions that are not
  finite-supported, the problem becomes significantly more
  complicated: the ``dominant ancestor'' trait will keep changing with
  time.
  
\item
  According to~\eq{Modd}, we have $\skew=\O{\t^{-9/2}}$, so the
  related correction $\frac12\fitg\stddev^2\skew$ in~\ref{C-var} is
  $\O{\t^{-7/2}}$, which is asymptotically smaller than the main term
  $\mut/\stddev$, which is $\O{t^{-1/2}}$.
\end{itemize}

\end{document}